\documentclass[sigconf]{acmart}
\usepackage{pifont}
\usepackage{natbib} 
\usepackage{booktabs}
\usepackage{comment}
\usepackage{graphicx}
\usepackage{subfig}
\usepackage{fancyhdr}
\usepackage{amsmath}
\usepackage{xurl}
\usepackage{adjustbox}
\usepackage{arydshln}
\usepackage{relsize}
\usepackage{wasysym}
\usepackage{multirow}
\usepackage[makeroom]{cancel}
\usepackage{scalerel,graphicx,xparse}
\usepackage{algorithm}
\usepackage{algpseudocode}
\usepackage[capitalize,nameinlink]{cleveref}

\usepackage{lscape}
\usepackage{graphicx}
\usepackage{enumitem}

\graphicspath{{../}{images/}}

\definecolor{navy}{rgb}{0.1, 0.1, 0.8}
\definecolor{gray}{rgb}{0.4, 0.4, 0.4}
\definecolor{olive}{rgb}{0.1, 0.5, 0.1}
\definecolor{ruby}{rgb}{0.8, 0.1, 0.3}
\definecolor{darkpastelgreen}{rgb}{0.01, 0.75, 0.24}
\definecolor{celestialblue}{rgb}{0.29, 0.59, 0.82}
\definecolor{coral}{rgb}{1.0, 0.5, 0.31}
\definecolor{blue}{rgb}{0.23, 0.44, 0.62}
\definecolor{Goldenrod}{rgb}{0.8,0.8,0}
\definecolor{pinky}{RGB}{255,20,147}

\usepackage{soul}
\usepackage[colorinlistoftodos,prependcaption,textsize=tiny,textwidth=15mm]{todonotes}
\usepackage{xspace}

\newcommand{\eat}[1]{}

\usepackage{listings}

\lstdefinelanguage{prompt}{
  sensitive=false,
  morecomment=[l]{\#},
  morestring=[b]",
}

\lstdefinestyle{promptstyle}{
  language=prompt,
  basicstyle=\ttfamily\small,
  breaklines=true,
  frame=single,
  backgroundcolor=\color{gray!5},
  captionpos=b,
  xleftmargin=0.5em,
  xrightmargin=0.5em,
  columns=fullflexible
}
\AtBeginDocument{%
  }

\copyrightyear{2026}
\acmYear{2026}
\setcopyright{cc}
\setcctype{by}
\acmConference[WWW '26]{Proceedings of the ACM Web Conference 2026}{April 13--17, 2026}{Dubai, United Arab Emirates}
\acmBooktitle{Proceedings of the ACM Web Conference 2026 (WWW '26), April 13--17, 2026, Dubai, United Arab Emirates}
\acmPrice{}
\acmDOI{10.1145/3774904.3792681}
\acmISBN{979-8-4007-2307-0/2026/04}

\begin{document}

\title{\textsc{Dreams}: A Social Exchange Theory-Informed Modeling of Misinformation Engagement on Social Media}

\author{Lin Tian}
\affiliation{%
  \institution{University of Technology Sydney}
  \department{Behavioral Data Science} 
  \city{Sydney}
  \country{Australia}}
\email{Lin.Tian-3@uts.edu.au}

\author{Marian-Andrei Rizoiu}
\affiliation{%
  \institution{University of Technology Sydney}
  \department{Behavioral Data Science} 
  \city{Sydney}
  \country{Australia}}
\email{Marian-Andrei.Rizoiu@uts.edu.au}


\begin{abstract}
Social media engagement prediction is a central challenge in computational social science, particularly for understanding how users interact with misinformation. 
Existing approaches often treat engagement as a homogeneous time-series signal, overlooking the heterogeneous social mechanisms and platform designs that shape how misinformation spreads.
In this work, we ask: ``Can neural architectures discover social exchange principles from behavioral data alone?'' 
We introduce \textsc{Dreams} (\underline{D}isentangled \underline{R}epresentations and \underline{E}pisodic \underline{A}daptive \underline{M}odeling for \underline{S}ocial media misinformation engagements), a social exchange theory-guided framework that models misinformation engagement as a dynamic process of social exchange. 
Rather than treating engagement as a static outcome, \textsc{Dreams} models it as a sequence-to-sequence adaptation problem, where each action reflects an evolving negotiation between user effort and social reward conditioned by platform context. 
It integrates adaptive mechanisms to learn how emotional and contextual signals propagate through time and across platforms. 
 On a cross-platform dataset spanning $7$ platforms and 2.37M posts collected between 2021 and 2025, \textsc{Dreams} achieves state-of-the-art performance in predicting misinformation engagements, reaching a mean absolute percentage error of $19.25$\%. 
 This is a $43.6$\% improvement over the strongest baseline. 
 Beyond predictive gains, the model reveals consistent cross-platform patterns that align with social exchange principles, suggesting that integrating behavioral theory can enhance empirical modeling of online misinformation engagement. 
 The source code is available at: \url{https://github.com/ltian678/DREAMS}.
\end{abstract}




\maketitle

\begin{figure}[t]
\centering
  \includegraphics[width=\linewidth]{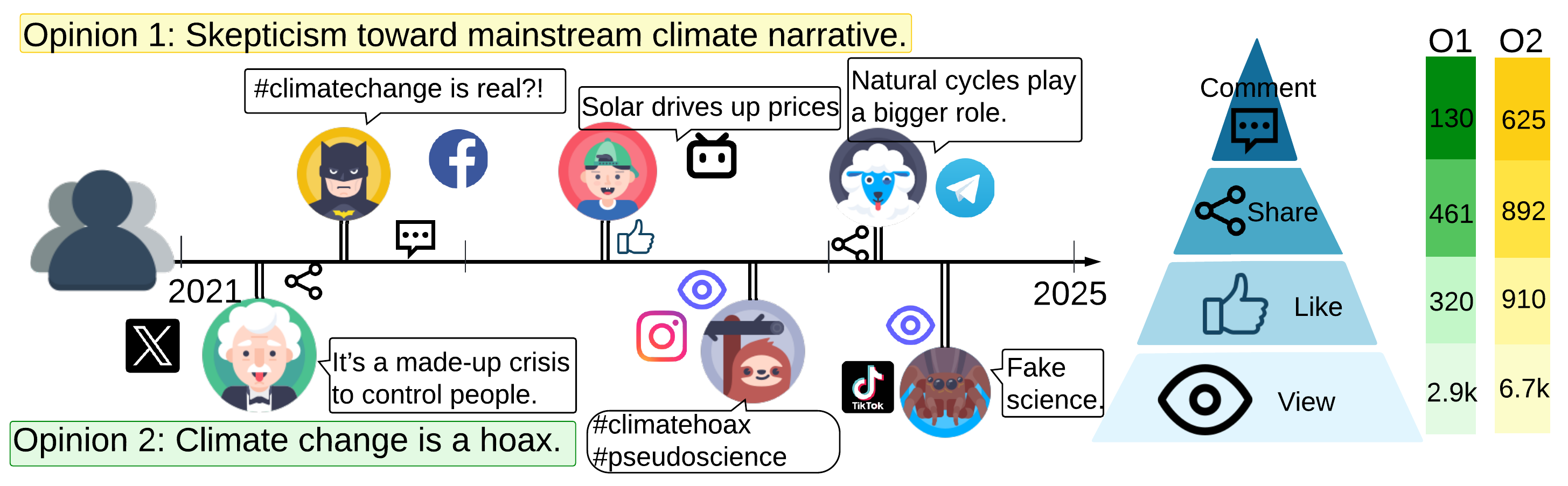}
  \caption{Overview of social media misinformation engagement dynamics and the proposed engagement pyramid.
}
  \label{fig:background}
\end{figure}
\section{Introduction}
What drives the obsessive engagement of over 5.4 billion people, who in 2025 spend an average of 2 hours and 24 minutes daily on platforms like Instagram, Facebook, and X (formerly Twitter)?\footnote{\url{https://datareportal.com/reports/digital-2025-global-overview}}
These interactions, such as scrolling, liking, sharing, and commenting, form a structured hierarchy of effort that we term the \textit{engagement pyramid} (\cref{fig:background}): from passive viewing at the base to active commenting at the apex.
Each step upward reflects a deeper social investment---time, cognition, and emotion---exchanged for visibility, recognition, or connection.

Social Exchange Theory (SET), which views social behavior as a negotiated exchange of costs and rewards, provides a conceptual foundation for understanding engagement on social media platforms.
In traditional social settings, SET explains how individuals seek to maximize social rewards such as approval or recognition while minimizing effort or risk~\citep{homans1958social,thibaut1959social,emerson1976social}.  
We hypothesize that online platforms preserve the core principles of Social Exchange Theory but modify their mechanisms: platform design and audience structure redefine what constitutes ``cost'' (e.g.,\ time, emotional expression) and ``reward'' (e.g.,\ likes, shares, comments).

Misinformation---information that is false or misleading regardless of intent---spread through social media's attention economy by generating strong emotional responses and engagement~\cite{budak2024misunderstanding,Booth2024}.
Polarizing content often spreads faster and engages more users, highlighting how platform structures triger emotional responses~\cite{vosoughi2018spread,ecker2022psychological,Lee2024}.
Understanding engagement with misinformation, therefore, is not merely a predictive task but a behavioral one: it requires modeling how users invest attention and emotion in response to content that manipulates perceived social value.

\begin{figure*}[t]
\centering
  \includegraphics[width=0.9\linewidth]{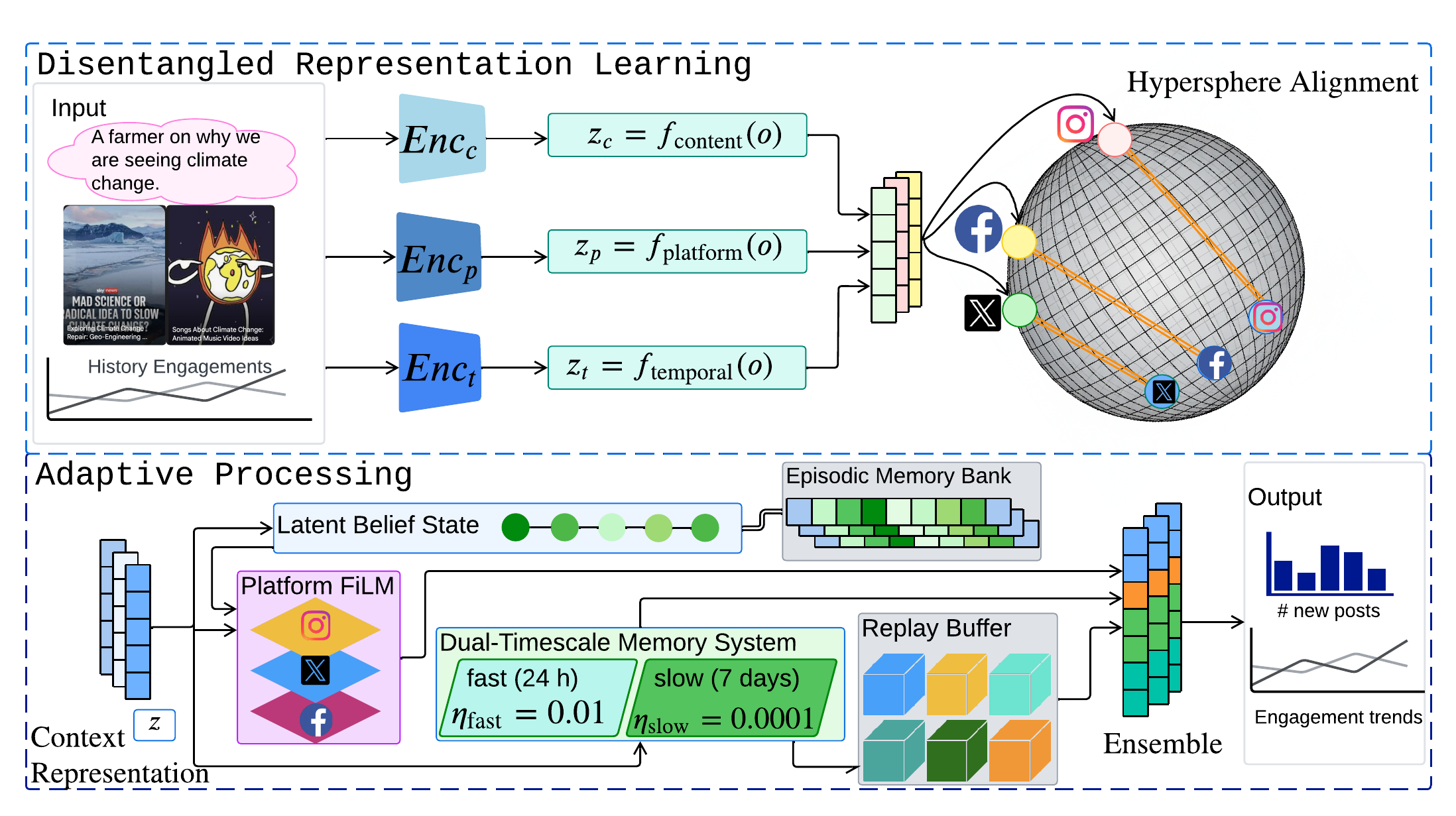}
  \caption{Overview of the \textsc{Dreams} architecture for cross-platform misinformation engagement prediction. 
  \textit{(Top)} The encoder disentangles input features into three latent factors (content ($z_c$), platform ($z_p$), temporal dynamics ($z_t$)), which then projected onto a hypersphere for scale-invariant representations.
  \textit{(Bottom)} The aligned representations feed into four SET-driven components: 
  (i) a Latent Belief State to track exchange context, 
  (ii) Dynamic Architecture Adaptation via FiLM layers with sparse gating, 
  (iii) a Dual-Timescale Memory System with fast and slow adapters, 
  and (iv) a Replay Buffer to retain past engagement patterns. 
  The dual-head ensemble combines these components to jointly predict misinformation engagement levels and the number of new posts per platform-topic pair.}
  \label{fig:dreams_archi}
\end{figure*}

Early computational studies have explored engagement prediction through user activity, content features, and network structure~\citep{peters2024context,arazzi2023predicting,li2024delving}, yet these approaches treat engagement as a static outcome rather than as a dynamic exchange process. 
They overlook how users adapt future actions based on previous feedback, or how platform architectures transform the perceived value of engagement itself. 
In this work, we model misinformation engagement as a platform-conditioned exchange process between contextual investment and social return.

\textbf{Our contributions:}  
\textsc{Dreams}, an adaptive sequence-to-sequence framework for predicting misinformation engagement, making the following two key contributions:

\textit{Theory-informed integration of SET with sequence modeling.}  
We formalize engagement as a platform-conditioned exchange process between emotional investment and social return, combining disentangled representations, FiLM-based platform adaptation, and dual-timescale memory to capture both global reciprocity patterns and platform-specific exchange dynamics.

\textit{A large-scale, cross-platform dataset.} 
We construct a fine-grained, multimodal, and multilingual dataset for misinformation engagement prediction across $7$ social media platforms (2021--2025), allowing comparison of engagement behaviors across times, languages, and platforms.

\section{Related Work}
\textit{Modeling Social Media Misinformation Engagement.}  
Predicting engagement links content, audience, and platform dynamics~\citep{peters2024context,li2024delving,arazzi2023predicting,immorlica2024clickbait}.  
Early work used feature-based regression on posts popularity prediction~\citep{li2017deepcas}, while deep sequence models later treated engagement as temporal point processes shaped by prior posts and reactions~\citep{rizoiu2017expecting,rizoiu2022interval,kong2023interval,Calderon2025}.  
\citet{Calderon2024a} explored how competitive and cooperative opinion dynamics shape the information spread
and the interplay between content, source credibility, and sharing pathways \cite{Calderon2024b}. 
Most remain platform-specific and purely predictive, limiting generalization across heterogeneous environments.  
\textsc{Dreams} extends this by embedding SET, viewing engagement as a dynamic exchange between effort and reward.

\textit{Theory-Guided Behavioral Modeling.}  
Recent studies incorporate domain theory into neural systems to align learning with behavioral dynamics~\citep{prantikos2023physics,escapil2023h,Yuan2025}.  
In social contexts, psychological constructs such as reciprocity and trust have been modeled through reinforcement or state-space formulations~\citep{mahmoodi2018reciprocity,dao2022flashattention,Ram2024},
but large-scale engagement prediction rarely leverages such theoretical grounding.  
\textsc{Dreams} operationalizes SET within a sequence-to-sequence framework, linking latent representations to the exchange principles.

\textit{Cross-Platform Prediction.} 
Cross-platform and multimodal transfer methods have been widely explored in domains such as computer vision, information retrieval task, and content creation~\citep{wan2024cross,cheng2024xrag,ji2024vexkd}.  
However, their applications to misinformation engagement prediction remains limited, as existing models typically treat platforms as homogeneous channels \cite{Kong2023} and overlook the distinct social mechanisms that govern interaction.  
\textsc{Dreams} models platform-specific exchange dynamics through adaptive processing, providing a unified yet flexible representation of social behavior across multiple environments for misinformation engagement prediction.

\section{Formalizing SET for Engagement Modeling}
We formalize the core principles of SET into mathematical constructs that guide the design of \textsc{Dreams}.  
Social media interactions---clicks, likes, shares, and comments---are viewed as economic exchanges in which users invest attention and effort in expectation of social return.  
At each time step $t$, an interaction is represented as a tuple $(I_t, R_t, p_t, c_t)$, where $I_t$ denotes investment (time, attention, cognitive effort), $R_t$ denotes return (social rewards such as likes or replies), and $(p_t, c_t)$ are the platform and content context.  
The instantaneous utility is defined as
$
U_t = R_t - I_t \cdot \kappa_{p_t},
$
where $\kappa_{p_t}$ is the platform-specific interaction cost.  
We then introduce three hypotheses that motivate the architecture of \textsc{Dreams}.

\textbf{Hypothesis 1: Platform-Specific Exchange Rates.}
Different platforms have different costs for similar actions due to variations in interface design, audience expectations, and engagement mechanisms~\cite{emerson1976social}. We formalize this by proposing that for the same content $c$, the exchange function varies by platform according to $f_p: \text{Effort} \mapsto \text{Reward}$, where $f_p(e) = \gamma_p \cdot g(e) + \beta_p$. Here, $\gamma_p$ represents platform-specific feature and $\beta_p$ captures baseline engagement.
In \textsc{Dreams}, this mechanism is applied through FiLM layers that apply feature-wise linear modulation conditioned on platform identity.  
Each $(\gamma_p, \beta_p)$ pair acts as a learned \textit{exchange rate}, adjusting the shared representation to reflect how each platform converts expressed opinions into engagement outcomes.

\textbf{Hypothesis 2: Reciprocity and Exchange Memory.}
Past exchanges create expectations that influence future behavior, as the norm of reciprocity~\citep{gouldner1960norm, blau1964}. We frame this relationship by proposing that future engagement depends on exchange history as:\\
$
\mathbb{E}[R_{t+k}|H_t] = \psi\left(\sum_{\tau=1}^{T} w_\tau \cdot (R_{t-\tau} - I_{t-\tau})\right)
$
where $H_t$ is the history up to time $t$, and $w_\tau$ is temporal decay of reciprocal obligations.
To capture both immediate and delayed reciprocity, \textsc{Dreams} uses a dual-timescale memory system: a fast memory operating on $\sim$24-hour windows for short-term exchanges, and a slow memory spanning $\sim$7-day windows for long-term relational dynamics.  

\textbf{Hypothesis 3: Emotional Currency in Platform-Specific Exchange Markets.}
Emotions act as social currency in online interactions, where engagement measures the reciprocal value of emotional expression~\cite{berger2012makes,frijda1986emotions}.  
Their impact depends on how each platform transforms emotional signals into user response. 
We model this process as
$
V_{\text{exchange}} = \alpha\, E_{\text{emotion}}\, \Phi_{\text{platform}} + \beta\, C_{\text{context}},
$
where $E_{\text{emotion}}$ is emotional intensity, $\Phi_{\text{platform}}$ the platform-specific conversion function, and $C_{\text{context}}$ contextual influence.  
Coefficients $\alpha$ and $\beta$ govern emotional and contextual contributions, treating each platform as a distinct market with its own conversion rate between emotion and engagement.  
To capture these heterogeneous exchange dynamics, \textsc{Dreams} uses a variational disentanglement framework to separate content value $z_c$ (\textit{what} is shared), platform context $z_p$ (\textit{where}), and temporal dynamics $z_t$ (\textit{when}).

\section{Methodology}
\begin{algorithm}[t]
\caption{\textsc{Dreams} Training Framework}
\label{alg:dreams_training}
\begin{algorithmic}[1]
\State \textbf{Input:} Training data $\mathcal{D}=\{(o_i,p_i,t_i,y_i,n_i)\}_{i=1}^N$
\State \textbf{Modules:} Encoders $\text{Enc}_c,\text{Enc}_p,\text{Enc}_t$; Context MLP; Backbone $\mathcal{B}_\theta$; Fast/Slow adapters $A_f,A_s$; Episodic memory $M_{\text{epi}}$; FiLM $\{\gamma_p,\beta_p\}$; Heads $f_{\text{last}},f_{\text{avg}},f_{\text{vol}}$; NF
\State \textbf{Buffers:} Replay buffer $\mathcal{B}$ (capacity $|\mathcal{B}|=10^4$)
\State \textbf{Output:} Trained parameters $\Theta$

\For{epoch $=1$ \textbf{to} $E$}
  \For{\textbf{each} batch $\{(o,p,t,y,n)\}\subset\mathcal{D}$}
    \vspace{2pt}
    \State \textit{// Disentangle \& projection on unit sphere}
    \State \label{line:enc} $z_c \gets \text{Enc}_c(o)$;\ \ $z_p \gets \text{Enc}_p(p)$;\ \ $z_t \gets \text{Enc}_t(t)$
    \State $z_{\text{ctx}} \gets \text{MLP}_\text{ctx}([z_c;z_p;z_t])$;\ \ $z \gets z_{\text{ctx}} / \|z_{\text{ctx}}\|_2$
    \State $\mathcal{L}_{\text{dis}} \gets \beta\, D_{\mathrm{KL}}\!\big(q(z|o,p,t)\,\|\,\mathcal{N}(0,I)\big) + \lambda_{\text{dis}} \sum_{i\neq j} I(z^i;z^j)$
    \State \textit{// Dual-timescale memory features}
    \State $m_f \gets A_f(\mathcal{X}_{[t-\tau_f,t]};\theta_f)$;\ \ $m_s \gets A_s(\mathcal{X}_{[t-\tau_s,t]};\theta_s)$
    \State $\alpha \gets \sigma(W_\alpha [z_c;z_t])$;\ \ $M \gets \alpha\, m_f + (1-\alpha)\, m_s$
    \vspace{3pt}
    \State \textit{// Backbone state and episodic retrieval}
    \State $s_t \gets \mathcal{B}_\theta(W_{\text{proj}} z, s_{t-1})$
    \State $r \gets \text{Attention}(s_t, M_{\text{epi}})$
    \State $b \gets \text{LayerNorm}(s_t + r)$;\ \ $b_{\text{flow}} \gets \mathrm{NF}(b)$
    \State $b_{\text{comp}} \gets b_{\text{flow}} - \mathrm{Mean}\big(\phi(\mathcal{N}_T)\big)$ 
    \vspace{3pt}
    \State \textit{// Platform adaptation (FiLM)}
    \State $h \gets \gamma_p \odot b_{\text{comp}} + \beta_p$
    \State $\hat{y} \gets w \cdot f_{\text{last}}([h;m_f;z_c]) + (1-w)\cdot f_{\text{avg}}([h;m_s;z_c])$ 
    \State $\hat{n} \gets f_{\text{vol}}([h;M;z_c])$ \Comment{new-post count}
    \vspace{3pt}
    \State \textit{// Losses \& update}
    \State $\mathcal{L}_{\text{pred}} \gets \mathrm{MAPE}(\hat{y},y) + \mathrm{MAPE}(\hat{n},n)$
    \State $\mathcal{L}_{rep} \gets \text{MAPE}(\text{DREAMS}(\text{Sample}(\mathcal{B})), y_\text{rep})$ | $|\mathcal{B}| > 0$
    \State $\mathcal{L} \gets \mathcal{L}_{\text{pred}} + \lambda_1 \mathcal{L}_{\text{dis}} + \lambda_2 \mathcal{L}_{\text{rep}}$
    \State $\Theta \gets \Theta - \eta \nabla_\Theta \mathcal{L}$ \Comment{update all trainable parameters}
    \vspace{3pt}
    \State \textit{// Memory write }
    \State $g_{\text{write}} \gets \sigma\!\big(W_g [z; M; e]\big)$ 
    \State $M_{\text{epi}} \gets \mathrm{Write}\big(s_t, M_{\text{epi}}, g_{\text{write}}\big)$
    \vspace{3pt}
    \State \textit{// Update replay buffer}
    \State $\mathcal{B} \gets \mathrm{UpdateBuffer}\big(\mathcal{B}, (o,p,t,y,n)\big)$
  \EndFor
\EndFor
\State \Return $\Theta$
\end{algorithmic}
\end{algorithm}

This section defines the misinformation engagement prediction task, introduces the \textsc{Dreams} architecture for cross-platform modeling, and explains the training objectives for model learning.

\subsection{Problem Formulation}
We frame cross-platform misinformation engagement prediction on social media as a multi-task sequence learning problem at the \textit{opinion level}. 
Let $\mathcal{P} = \{p_1, p_2, ..., p_P\}$ denote the set of platforms and $\mathcal{T} = \{t_1, t_2, ..., t_T\}$ the discrete time steps. 
At each time $t$, we observe a collection posts $\mathcal{X}_t = \{x_i\}_{i=1}^{N_t}$ expressing opinion $o$, where each post $x_i = (c_i, p_i, h_i)$ has content features $c_i \in \mathbb{R}^{d_c}$ (text, image, or video), a platform identifier $p_i \in \mathcal{P}$, and its engagement history $h_i = \{e_i^{(\ell)}\}_{\ell=0}^{L}$. 
We define engagement as a hierarchy of four user actions with increasing effort: $\mathcal{E} =\{e^{(0)}:\tiny{View}, \ e^{(1)}:\tiny{Like}, \ e^{(2)}:\tiny{Share}, \ e^{(3)}:\tiny{Comment}\}$. 
Each level represents progressively higher cognitive and temporal investment.

Given data up to time $t$, \textsc{Dreams} jointly predicts:  
(1) engagement trajectories $\hat{y}_o^{(\ell)}$ for each opinion $o$ and engagement level $\ell$ at a future time $t+\tau$, and  
(2) the expected number of new posts $\hat{n}_p$ per platform $p$ at the same time step. 

\subsection{\textsc{Dreams} Architecture}
\textsc{Dreams} is architecture-agnostic and can augment any sequence-to-sequence model (e.g.,\ Transformer, Mamba or LSTM). In our experiments, we implement two variants: D-Mamba and D-Transformer, corresponding to Mamba- and Transformer-based backbones, respectively.
The full training procedure is described in \cref{alg:dreams_training}.

\textsc{Dreams} operates through two key processes, illustrated in \cref{fig:dreams_archi}:  
(1) \textit{Disentangled representation learning}, which separates content, platform, and temporal influences into independent latent variables; and  
(2) \textit{Adaptive processing}, which enables the sequential model to adjust dynamically across platforms and timescales.


\subsubsection{Disentangled Representations}
\label{subsec:distangle_rep}
%
Given an input $x_t$ and platform $p_t$, \textsc{Dreams} learns three disentangled latent factors ($z_c$, $z_p$,and $z_t$), as described in line 8 of \cref{alg:dreams_training}.
The disentangled factors are combined through a lite context module ($\text{MLP}_\text{ctx}$) and projected onto a unit hypersphere (line 9 in \cref{alg:dreams_training}) to learn the aligned representation $z$.
This projection constrains all representations to locate on the unit sphere $\mathcal{S}^{d-1} \subset \mathbb{R}^d$ with $\|z\|_2 = 1$. It ensures scale invariance across platforms, where engagement is compared by angle rather than magnitude, and maintains gradient stability within a bounded representation space.
To further separate the latent factors, we apply a variational loss with a mutual information penalty (line 10 in \cref{alg:dreams_training}), where $\beta$ and $\gamma_{\text{dis}}$ control the strength of the Gaussian prior $p(z)=\mathcal{N}(0,I)$.

\subsubsection{Latent Belief State}
\textsc{Dreams} includes a \textit{latent belief state} that combines the model's current observation with past episodic context.
Rather than performing explicit POMDP-style updates, it encodes a probabilistic summary of how the current interaction relates to previous engagement patterns.
Given the disentangled and projected representation $z$, the backbone produces a hidden state $s$, which retrieves relevant past episodes $r$ from the episodic memory $M_{\text{epi}}$ through attention (lines 15--16 in \cref{alg:dreams_training}).
The latent belief state $b$ is then formed by fusing both signals using a residual connection and normalization (line 17 in \cref{alg:dreams_training}).
To ensure $b$ keeps a valid belief distribution, we apply a normalizing flow $\text{NF}(\cdot)$ to map the fused representation to a probabilistically valid space, followed by a comparison mechanism that subtracts the mean of recent neighborhood exchanges $\mathcal{N}_T$ within temporal window $T$ (line 18 in \cref{alg:dreams_training}). When $b_{\text{comp}} > 0$, the current opportunity exceeds historical expectations, indicating higher engagement likelihood. The resulting $b_{\text{comp}}$ thus serves as a context-aware belief embedding that conditions subsequent FiLM modulation and prediction heads (\cref{subsec:film}).

\subsubsection{Dynamic Architecture Adaptation}
\label{subsec:film}
To adapt to the heterogeneous environments, \textsc{Dreams} uses \textit{dynamic architecture adaptation} through feature-wise modulation and attention gating.

\textbf{FiLM Modulation.} We apply Feature-wise Linear Modulation (FiLM)~\cite{perez2018film} to condition the model on platform context without duplicating parameters.
As shown in Line 15 of \cref{alg:dreams_training}, each platform $p$ has learned affine modulation parameters $(\gamma_p, \beta_p)$ that rescale and shift the belief embedding $b_{\text{comp}}$. This aims to normalize engagement magnitude across platforms while capturing the platform-specific differences in user behaviors. 

\textbf{Dynamic Attention Heads.} 
To further adapt to contextual feature dimensions per platform, we extend attention computation~\cite{vaswani2017attention} with platform-conditioned gating, as
$
\text{Attention}_{\text{adapt}} = \sum_{i=1}^K g_i \cdot \text{Head}_i(Q, K, V), \quad g_i = \sigma(W_{\text{head}} \cdot [h; p]).
$
The gating weights $g_i$ are functions of $h$ and $p$.

\textbf{Global Write Gate.} 
\textsc{Dreams} uses a \textit{global write gate} to control when new memory is added to the episodic memory $M_{\text{epi}}$.
The gate decides whether the current interaction should be stored as a new memory entry (line 29 \cref{alg:dreams_training}), using the aligned representation $z$, the combined fast-slow memory state $M$, and the observed engagement signal $e$.
It bridges short-term adaptation with long-term episodic retention.

\subsubsection{Dual-Timescale Memory System}
\label{subsec:memory_sys}
Social exchanges occur at different rhythms. 
Some are immediate (like-for-like within hours), while others unfold slowly (community interactions over weeks).
To model both patterns, \textsc{Dreams} uses a dual-memory architecture with three parts: (1) \textit{Fast Adapter} for short-term reciprocity, (2) \textit{Slow Adapter} for long-term relationship dynamics, and (3) \textit{Episodic Memory Bank} for storing episodes.

\textbf{Fast and Slow Memory.} 
\textsc{Dreams} uses two memory adapters to capture engagement dynamics at different timescales.
The fast adapter $m_f$ operates over a short time window($\tau_f = 24$h) with a higher learning rate ($\eta_f = 0.01$) to follow immediate trends. 
The slow adapter $m_s$ aggregates information over a longer period ($\tau_s = 7$ days) with a smaller learning rate ($\eta_s = 0.0001$) to capture stable patterns.
An adaptive gate then fuses the two memories to balance short-term responsiveness with long-term consistency (lines 12--13 in \cref{alg:dreams_training}).

\textbf{Episodic Memory Bank.} We use an external episodic memory $M_{\text{epi}} \in \mathbb{R}^{N \times d_m}$ storing $N$ prototypical exchange episodes. 
During inference, the backbone state $s_t$ attends to stored episodes to retrieve relevant historical exchanges (line 30 in \cref{alg:dreams_training}).
Here $\psi(s)$ produces a candidate key from the current state and $g_{\text{write}} = \sigma(W_g[z; M; e])$ is the global write gate controlling whether a new episode should be stored.

\textbf{Replay Buffer.} We implement a replay buffer $\mathcal{B}$~\cite{rolnick2019experience} to store up to $|\mathcal{B}| = 10^4$ past samples (line 32 in \cref{alg:dreams_training}). During training, \textsc{Dreams} periodically replays these examples to retain past engagement patterns while adapting to newer ones.

\subsubsection{Dual-head Prediction Ensemble}
\textsc{Dreams} integrates information from multiple temporal scales through a dual-head prediction ensemble.
The recency head $f_{\text{last}}$ focuses on recent patterns by using fast memory $m_f$. The $f_{\text{avg}}$ predictor averages over a longer history through slow memory $m_s$.  
Both are implemented as independent MLP predictors, and their outputs are combined using a learnable weight $w \in [0,1]$ jointly optimized with all model parameters (line 21 in \cref{alg:dreams_training}).

\subsection{Training Objectives}
\textsc{Dreams} is trained using three loss functions that guide representation learning, temporal modeling, and cross-platform adaptation.

\textbf{Prediction Loss.} The supervised loss combines mean absolute percentage error (MAPE) for engagement trend forecasting across four hierarchical levels and for discrete post-volume prediction (line 22 in \cref{alg:dreams_training}).

\textbf{Disentanglement Loss.} To enforce statistical independence between content, platform, and temporal factors, we apply a $\beta$-VAE objective~\cite{higgins2017beta} with mutual information penalty (line 10 in \cref{alg:dreams_training}), where $\beta$ controls the strength of the prior constraint and $\lambda_{\text{dis}}$ minimizes mutual information between latent dimensions to encourage factorized representations. 
Following~\cite{kim2018disentangling}, we estimate $I(z^i; z^j)$ via total correlation decomposition.

\textbf{Replay Loss.} To keep \textsc{Dreams} stable when learning from multi-year social data (2021--2025), we add a replay loss that helps the model remember past patterns while adapting to new ones (line 25 in \cref{alg:dreams_training}).
Following continual learning approach~\cite{rebuffi2017icarl}, a replay buffer $\mathcal{B}$ is used to store examples from earlier time periods.
During training, batches are composed of both new and replayed instances, and the model minimizes prediction error on both sets.

\section{Experiments and Results}
In this section, we first introduce a large-scale, multi-platform misinformation dataset collected between 2021 and 2025, covering diverse languages, topics, and platforms.
The dataset captures how users engage with both verified and misleading opinions.
We then evaluate \textsc{Dreams} against statistical, recurrent, and state space baselines to address three key research questions:
(RQ1) how platform-specific exchange rates shape engagement patterns;
(RQ2) how reciprocity and exchange memory influence user responses over time; and
(RQ3) how emotional currency modulates engagement dynamics across misinformation narratives. 
We also conduct ablation studies to analyze the contribution of each component.

\subsection{Dataset}
We evaluate on cross-platform engagement data from 2021--2025 spanning 7 platforms (X, Bluesky, TikTok, Bilibili, Instagram, Telegram, Facebook). 
The data are segmented into 30-minute bins with 7-day input windows for prediction at 7-day and 28-day horizons. 
All experiments use chronological splits to ensure temporal validity and prevent information leakage.
\cref{tab:dataset} summarizes dataset statistics, platform coverage, and engagement level mapping.
Additional dataset construction and preprocessing details are provided in Appendix \cref{app:dataset_more}.
\begin{table}[t]
\centering
\caption{Dataset statistics and engagement level availability across platforms. Level 0: Views/Plays, Level 1: Likes/Favourites/Emojis, Level 2: Shares/Retweets/Reposts, Level 3: Comments/Replies. \checkmark as available data, \ding{55} unavailable.}
\label{tab:dataset}
\vspace{-0.5\baselineskip}
\begin{tabular}{lrrcccc}
\toprule
\textbf{Platform} & \textbf{\# Posts} & \textbf{\# Users} & \textbf{L0} & \textbf{L1} & \textbf{L2} & \textbf{L3} \\
\midrule
Facebook & 1,997,375 & 42,381 & \ding{55} & \checkmark & \checkmark & \checkmark \\
Instagram & 273,563 & 8,926 & \ding{55} & \checkmark & \ding{55} & \checkmark \\
X & 45,446 & 3,218 & \ding{55} & \ding{55} & \checkmark & \checkmark \\
Telegram & 33,310 & 1,847 & \checkmark & \ding{55} & \checkmark & \ding{55} \\
Bilibili & 9,845 & 892 & \checkmark & \checkmark & \checkmark & \checkmark \\
Bluesky & 4,158 & 521 & \ding{55} & \ding{55} & \checkmark & \checkmark \\
TikTok & 3,960 & 763 & \checkmark & \checkmark & \checkmark & \ding{55} \\
\midrule
\textbf{Total} & \textbf{2,367,657} & \textbf{58,548} & -- & -- & -- & -- \\
\bottomrule
\end{tabular}
\end{table}
\begin{table*}[t]
\centering
\caption{Engagement prediction performance (MAPE) across platforms and time horizons. Lower is better. Best per platform \textbf{bold}, second best \underline{underlined}. Platform abbreviations: F = Facebook, I = Instagram, T = Telegram, B = Bilibili, S = Bluesky, K = TikTok, dis. = disentangled representations.}
\vspace{-1.0\baselineskip}
\label{tab:detailed_results}
\begin{tabular}{lc@{\;\;}c@{\;\;}c@{\;\;}c@{\;\;}c@{\;\;}c@{\;\;}c|c@{\;\;}c@{\;\;}c@{\;\;}c@{\;\;}c@{\;\;}c@{\;\;}c}
\toprule
\multirow{2}{*}{\textbf{Model}} & \multicolumn{7}{c|}{\textbf{7 Days}} & \multicolumn{7}{c}{\textbf{28 Days}} \\
\cmidrule(lr){2-8} \cmidrule(lr){9-15}
& F & I & X & T & B & S & K & F & I & X & T & B & S & K \\
\midrule
HistAvg~\cite{hyndman2018forecasting}   & 0.635 & 0.704 & 0.697 & 0.746 & 0.794 & 0.739 & 0.732 & 1.095 & 1.214 & 1.202 & 1.286 & 1.369 & 1.274 & 1.262 \\
ARIMA~\cite{box2015time}     & 0.485 & 0.538 & 0.533 & 0.570 & 0.607 & 0.565 & 0.559 & 0.870 & 0.964 & 0.955 & 1.021 & 1.087 & 1.011 & 1.002 \\
\midrule
LSTM~\cite{hochreiter1997long}     & 0.415 & 0.460 & 0.456 & 0.487 & 0.519 & 0.483 & 0.478 & 0.787 & 0.872 & 0.864 & 0.924 & 0.983 & 0.915 & 0.906 \\
TCN~\cite{bai2018empirical}       & 0.378 & 0.419 & 0.415 & 0.444 & 0.472 & 0.440 & 0.435 & 0.725 & 0.804 & 0.796 & 0.851 & 0.906 & 0.843 & 0.835 \\
\midrule
RoBERTa~\cite{liu2019roberta}& 0.628 & 0.696 & 0.689 & 0.737 & 0.784 & 0.730 & 0.723 & 1.462 & 1.621 & 1.605 & 1.716 & 1.827 & 1.700 & 1.684 \\
xLSTM~\cite{beck2024xlstm}    & 0.422 & 0.468 & 0.463 & 0.495 & 0.527 & 0.491 & 0.486 & 0.842 & 0.934 & 0.925 & 0.989 & 1.053 & 0.980 & 0.971 \\
\midrule
Informer~\cite{zhou2021informer}  & 0.363 & 0.403 & 0.399 & 0.426 & 0.454 & 0.422 & 0.419 & 0.735 & 0.815 & 0.807 & 0.863 & 0.919 & 0.855 & 0.847 \\
TS-Transf.~\cite{zerveas2021transformer} & 0.378 & 0.419 & 0.415 & 0.443 & 0.472 & 0.439 & 0.435 & 0.754 & 0.836 & 0.828 & 0.885 & 0.943 & 0.877 & 0.869 \\
Mamba~\cite{gu2024mamba}     & 0.339 & 0.376 & 0.372 & 0.398 & 0.424 & 0.395 & 0.391 & 0.704 & 0.781 & 0.773 & 0.827 & 0.880 & 0.819 & 0.811 \\
IC-Mamba~\cite{tian2025before}  & 0.283 & 0.333 & 0.330 & 0.353 & 0.376 & 0.350 & 0.346 & 0.595 & 0.708 & 0.701 & 0.749 & 0.798 & 0.742 & 0.735 \\
\midrule
\textbf{D-Mamba} & \textbf{0.170} & \textbf{0.188} & \textbf{0.186} & \textbf{0.199} & \textbf{0.212} & \textbf{0.197} & \textbf{0.195} & \textbf{0.395} & \textbf{0.437} & \textbf{0.433} & \textbf{0.463} & \textbf{0.493} & \textbf{0.459} & \textbf{0.455} \\
\ \ \ w/o FiLM & 0.201 & 0.223 & 0.221 & 0.236 & 0.251 & 0.234 & 0.231 & 0.474 & 0.525 & 0.520 & 0.556 & 0.592 & 0.551 & 0.546 \\
\ \ \ w/o dis. & 0.237 & 0.263 & 0.260 & 0.278 & 0.296 & 0.275 & 0.273 & 0.571 & 0.633 & 0.626 & 0.670 & 0.713 & 0.664 & 0.657 \\
\ \ \ w/o belief & 0.181 & 0.200 & 0.198 & 0.210 & 0.223 & 0.206 & 0.204 & 0.434 & 0.482 & 0.477 & 0.510 & 0.543 & 0.505 & 0.500 \\
\ \ \ w/o memory & 0.188 & 0.208 & 0.206 & 0.220 & 0.235 & 0.218 & 0.216 & 0.447 & 0.496 & 0.491 & 0.525 & 0.559 & 0.520 & 0.515 \\
\midrule
\textbf{D-Transf.} & \underline{0.202} & \underline{0.224} & \underline{0.222} & \underline{0.237} & \underline{0.253} & \underline{0.235} & \underline{0.233} & \underline{0.523} & \underline{0.580} & \underline{0.574} & \underline{0.614} & \underline{0.654} & \underline{0.608} & \underline{0.603} \\
\ \ \ w/o FiLM & 0.237 & 0.263 & 0.260 & 0.278 & 0.296 & 0.276 & 0.273 & 0.567 & 0.629 & 0.623 & 0.666 & 0.709 & 0.660 & 0.654 \\
\ \ \ w/o dis. & 0.272 & 0.302 & 0.299 & 0.319 & 0.340 & 0.316 & 0.313 & 0.588 & 0.652 & 0.646 & 0.691 & 0.735 & 0.684 & 0.678 \\
\ \ \ w/o belief & 0.262 & 0.291 & 0.288 & 0.308 & 0.328 & 0.305 & 0.302 & 0.557 & 0.618 & 0.612 & 0.654 & 0.697 & 0.648 & 0.642 \\
\ \ \ w/o memory & 0.224 & 0.249 & 0.246 & 0.263 & 0.280 & 0.260 & 0.258 & 0.550 & 0.609 & 0.604 & 0.645 & 0.687 & 0.639 & 0.633 \\
\bottomrule
\end{tabular}%
\end{table*}

\textit{Data Collection and Preprocessing.}
For X and Facebook, we use publicly available misinformation datasets with post-level engagement data~\citep{kong2022slipping}, covering 2020--2024.
Bluesky, Bilibili, Instagram, Telegram and TikTok data were collected via keyword-based crawling, and top daily hashtags to capture high-traffic topics. 
Video posts from Bilibili and TikTok were transcribed using automatic speech recognition for analysis\footnote{\url{https://openai.com/index/whisper/}}.  
Each post includes at least one engagement event after publication to ensure active visibility.  

\textit{Opinion Grouping and Labeling.}
Engagement prediction is conducted at the \textit{opinion level}, where posts expressing similar stances or narratives are grouped together.  
For Facebook, we use the original fine-grained misinformation opinion annotations~\cite{kong2022slipping}.  
For other platforms, we generate opinion clusters using a topic modeling pipeline that combines semantic embeddings, hierarchical clustering, and post-hoc narrative labeling (Appendix \cref{app:topic_label_pipeline}).
We apply a transfer learning approach by fine-tuning \texttt{LLaMA-3} model~\cite{dubey2024llama} (LoRA adaptation~\cite{hu2022lora}) on SocialSense dataset~\citep{kong2022slipping}, and extend it through in-context learning to label clusters from other platforms. 
This grouping consolidates individual posts into coherent opinion trajectories, allowing \textsc{Dreams} to model the collective engagement dynamics around shared narratives.

\textit{Dataset Scale and Contribution.}
The final dataset contains 2.36 million posts from 58.5K users across seven platforms, spanning diverse temporal, topical, and interactional contexts.  
It offers a unified, fine-grained, and temporally aligned view of cross-platform social engagement, enabling \textsc{Dreams} to learn how different platforms value and social exchange dynamics.

\subsection{Baselines}
We compare \textsc{Dreams} with statistical, recurrent, transformer-based, and state space forecasting models. 
Statistical baselines include Historical Average~\cite{hyndman2018forecasting} and ARIMA~\cite{box2015time}. 
Recurrent models include LSTM~\cite{hochreiter1997long} and xLSTM~\cite{beck2024xlstm}\footnote{\url{https://github.com/NX-AI/xlstm}}, the latter extending memory via exponential gating. 
Transformer-based models comprise Informer~\cite{zhou2021informer} and TS-Transformer~\cite{zerveas2021transformer} for long-sequence modeling, and a text-only RoBERTa baseline~\cite{liu2019roberta}. 
We also evaluate the Temporal Convolutional Network (TCN)~\cite{bai2018empirical}\footnote{\url{https://github.com/locuslab/TCN}} for causal convolutional forecasting, and state-space models Mamba~\cite{gu2024mamba} and IC-Mamba~\cite{tian2025before}, which incorporates interval-censored learning for sparse supervision.

%
%
%
%
%
%

\subsection{Misinformation Engagement Prediction}
The results in \cref{tab:detailed_results} show clear performance gaps across model families and platforms.  
Among traditional methods, ARIMA improves by 23.5\% over HistAvg, setting a strong baseline.
Generative models show mixed results: standard LSTM outperforms extended memory xLSTM ($0.468$ vs.\ $0.481$), and RoBERTa catastrophically fails ($0.703$), indicating that neither extended memory nor language pretraining benefits engagement prediction. 
Transformer variants achieve moderate gains, with Informer reaching $0.411$ via its sparse attention mechanism, as the best among attention-based models. 
State space models deliver the strongest results: vanilla Mamba ($0.385$) improves further with IC-Mamba ($0.341$) and peaks with D-Mamba ($0.19$).  
D-Mamba surpasses vanilla Mamba by 50.6\% and Informer by 53.8\%, showing that incorporating domain-specific biases could boost the performance.

\textbf{Across platforms, performance aligns with interaction complexity rather than content type.}
Facebook shows the lowest error ($0.17$), followed by Instagram ($0.188$) and X ($0.186$).
Platforms with more complex or layered interaction designs perform worse: Telegram ($0.199$), TikTok ($0.195$), and Bilibili ($0.212$).
All platforms show consistent decay from 7-day to 28-day forecasts, as the limits of long-term predictability.
The weak effect of long memory (xLSTM underperforming LSTM by 2.8\%) and the universal temporal decay suggest that engagements are approximately Markovian, which is dominated by recent context rather than extended history.
\cref{tab:new_posts_mape} shows the performance of \textsc{Dreams} on forcasting how many new posts appear across different platforms.
Prediction is easier on community-oriented platforms like Facebook and Bilibili, where posting behavior is more regular, but harder on fast-paced platforms like X and TikTok, where activity changes quickly in response to trends.
\begin{table}[t]
    \centering
    \small
    \caption{Post-volume prediction performance (MAPE) over 7 days across platforms.
        F = Facebook, I = Instagram, T = Telegram, B = Bilibili, S = Bluesky, K = TikTok, -m = remove memory, -b = remove belief.}
    \label{tab:new_posts_mape}
    \begin{tabular}{lccccccc}
        \toprule
        \textbf{Model} & F     & X     & I     & K     & T     & B     & S     \\
        \midrule
        D-Mamba        & 0.281 & 0.526 & 0.355 & 0.425 & 0.470 & 0.389 & 0.510 \\
        D-Transf.      & 0.330 & 0.615 & 0.407 & 0.485 & 0.564 & 0.453 & 0.589 \\
        D-Mamba-m      & 0.311 & 0.679 & 0.358 & 0.457 & 0.589 & 0.419 & 0.658 \\
        D-Transf.-m    & 0.397 & 0.812 & 0.483 & 0.597 & 0.672 & 0.548 & 0.707 \\
        D-Mamba-b      & 0.385 & 0.525 & 0.421 & 0.503 & 0.514 & 0.469 & 0.547 \\
        D-Transf.-b    & 0.422 & 0.651 & 0.468 & 0.559 & 0.558 & 0.508 & 0.593 \\
        \bottomrule
    \end{tabular}
\end{table}

\begin{figure}[t]
\centering
  \includegraphics[width=\linewidth]{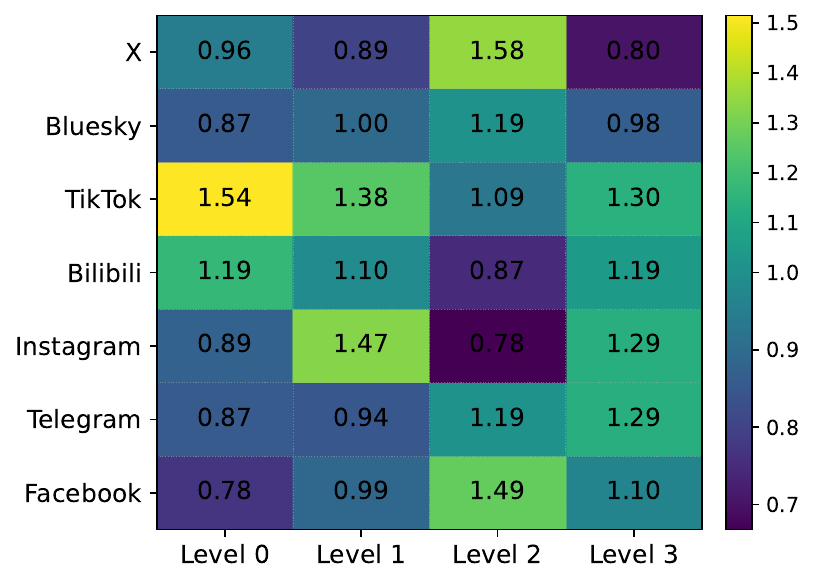}
  \caption{Platform-specific exchange rates learned by \textsc{Dreams}. Heatmap showing relative likelihood values ($\gamma$ parameters) extracted from FiLM layers for each platform-engagement level combination.}
  \label{fig:exchange_rate}
\end{figure}

\subsection{Platform-specific Exchange Rates}
To investigate how user engagement differs across platforms, we analyze the FiLM modulation parameters learned by \textsc{Dreams} for each platform and engagement level.
These parameters act as exchange rates, adjusting how much each platform amplifies or downplays a given social action during inference.

\textbf{Each platform values the same social actions differently.}
As shown in \cref{fig:exchange_rate}, \textsc{Dreams} learns these rates from user interaction patterns, showing how different platform designs assign value to various forms of engagement.
TikTok shows the highest value for level 0 (view) interactions ($\gamma$=$1.54$), reflecting its infinite scroll design that rewards passive consumption.
Because the algorithm optimizes for watch time, viewing becomes the main social currency, while higher engagement levels receive lower weights.
Instagram peaks at Level 1 (like) ($\gamma$=$1.47$), matching its visual-first design where the double-tap serves as the key signal of approval.
X shows relatively balanced exchange rates across levels, with heavier for shares (Level 2: $1.58$). 
This reflects the platform's retweet culture, where redistribution rather than direct reaction is the main form of value exchange~\cite{kwak2010twitter}.
However, Facebook shows an inverted patterns, with lower passive viewing (Level 0: $0.78$) but moderate comment reactions (Level 3: $1.10$).

\subsection{Reciprocity and Exchange Memory}
To understand how reciprocal actions occur at different timescales, we analyze the memory activation patterns learned by \textsc{Dreams}-Mamba. 
The model adjusts how much it relies on fast or slow memory when predicting engagement.
We extract fast and slow memory weights from correctly predicted test samples (MAPE < 10\%) and visualize their joint distribution.

\begin{figure}[t]
\centering
  \includegraphics[width=0.85\linewidth]{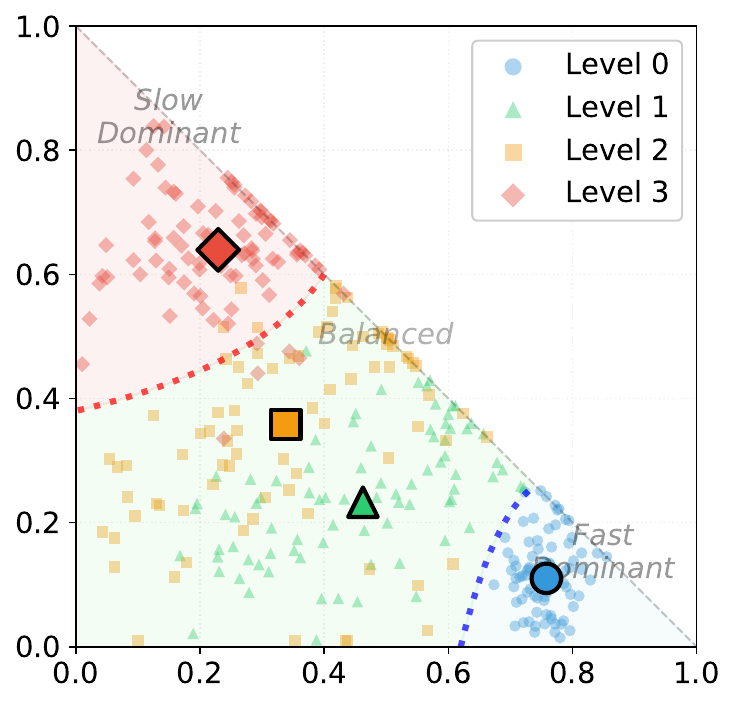}
  \caption{Fast versus slow memory activation weights from \textsc{Dreams}-Mamba for test instances with MAPE $<10$\%. Each point is a correctly predicted instance, colored by different engagement levels, with large markers indicating level-wise centroids. The memory weights are extracted from the dual-timescale memory banks after the forward pass through \textsc{Dreams}-Mamba.}
  \label{fig:memory}
\end{figure}

\textbf{Reciprocal engagement occurs at different timescales, with deeper interactions relying on longer memory.}
\cref{fig:memory} shows that how \textsc{Dreams} separates engagement types into three distinct memory regions, supporting our Hypothesis 2 that reciprocity operates at multiple temporal levels. 
level 0 (view) interactions appear in the fast-dominant region, reflecting passive consumption or ``lurking'' behaviors where users stay aware of others' content without direct exchange~\cite{lazer2009computational}. 
Interestingly, Level 1 (like) and Level 2 (share) is in overlapping regions in the balanced zone. 
These actions are low-effort and single-click movements---lightweight social signals~\cite{goffman1959presentation}---that provide recognition without building strong relational ties. 
It also likely reflects a platform design constraint: both are single-click actions requiring a similar amount of cognitive effort.
In contrast, Level 3 shows clear slow-memory dominance, as conversational engagements would require more effort.

The learned memory allocation aligns with network theories of tie strength~\cite{granovetter1973strength}. 
Weak ties (views and likes) operate through fast memory modules. Strong ties (comments) activate slow memory systems that encode relationship history and future obligations. This confirms our theoretical prediction that deeper engagements rely on longer memory traces.

%
\begin{figure*}[t]
    \centering
    \newcommand\myheight{0.167}

    \subfloat[]{
         \includegraphics[height=\myheight\textheight]{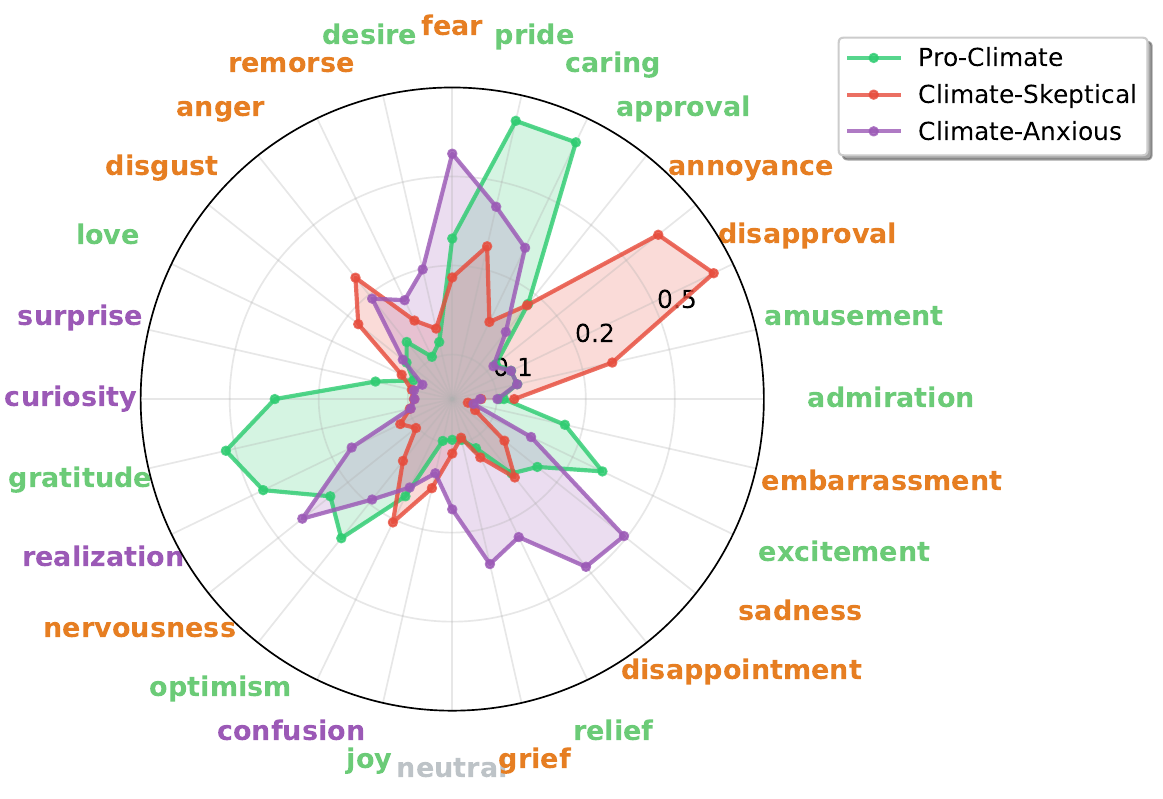}
         \label{fig:cc_emo}
    }%
    \subfloat[]{
        \includegraphics[height=\myheight\textheight]{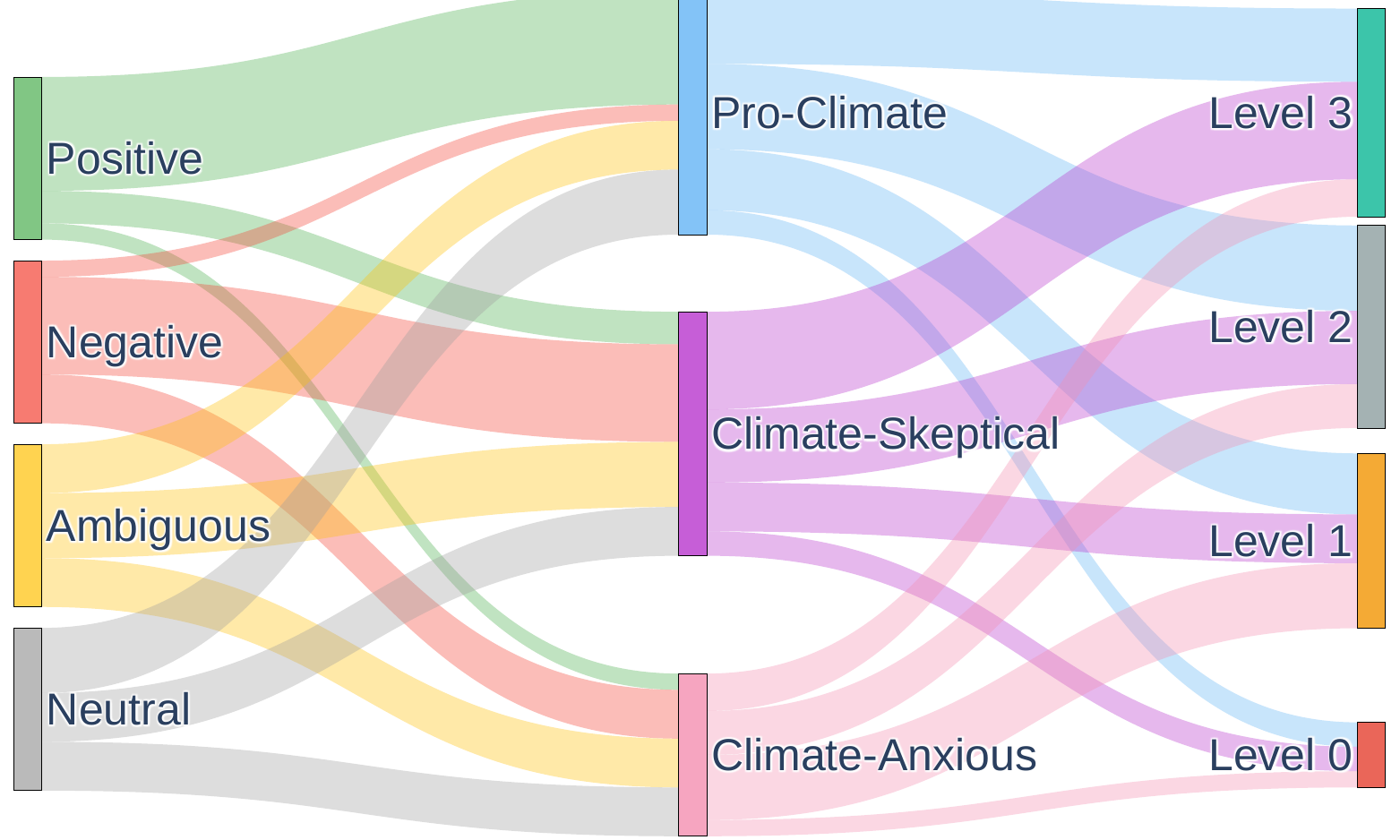}
        \label{fig:cc_emo_sankey}
    }%
    \subfloat[]{
        \includegraphics[height=\myheight\textheight]{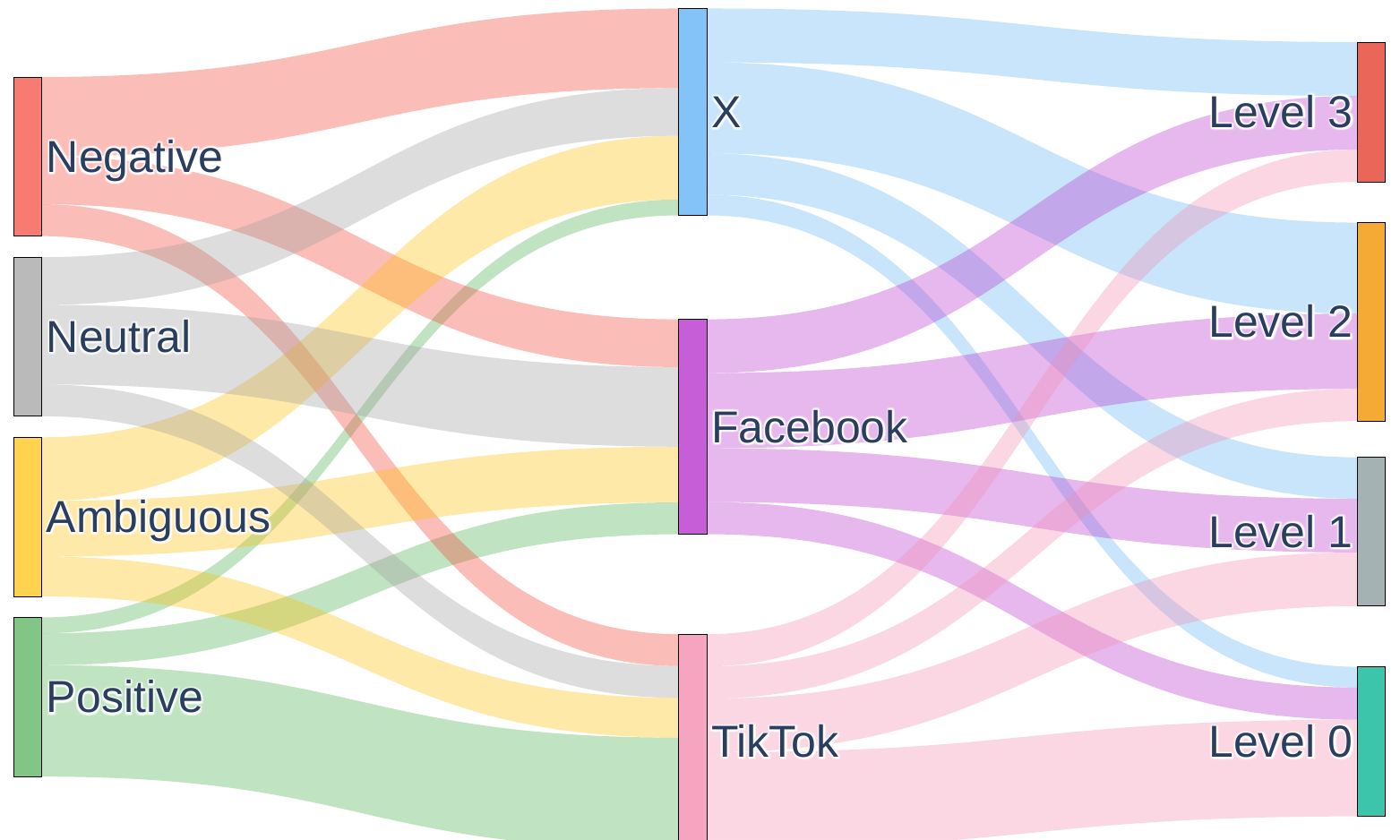}
        \label{fig:dog_emo_sankey}
    }%
    \caption{Emotional dynamics across topics, platforms, and \textsc{Dreams} predictions.
    (a) Emotional composition of three climate change narratives (\textit{Pro-Climate}, \textit{Climate-Skeptical}, and \textit{Climate-Anxious}) on Facebook (2022), across 28 GoEmotions categories (intensity $0.1$--$0.7$).
    (b) Emotional flows from aggregated GoEmotions categories through three climate change narratives to predicted engagement levels on \textit{Facebook}.
    (c) Emotional flows from aggregated GoEmotions categories through three platforms (X, Facebook, TikTok) to engagement levels for the \textit{dog adoption} topic (2022).
    }
    \label{fig:rq3_emotion_fig}
\end{figure*}
\subsection{Emotional Currency}
We apply a pretrained model\footnote{\url{https://huggingface.co/bsingh/robertagoEmotion}} on the GoEmotions dataset~\cite{demszky2020goemotions} to extract fine-grained emotion labels from post content.
The resulting emotion distributions, shown in \cref{fig:cc_emo}, serve as inputs for the engagement analyses in \cref{fig:cc_emo_sankey}, where emotions are aggregated into four categories (\textit{positive}, \textit{negative}, \textit{ambiguous}, and \textit{neutral}) to quantify cross-platform and cross-narrative differences based on the mapping between emotion and predicted engagement.

\textbf{Each platform shapes emotional currency uniquely, defining how emotions translate into engagement.}
We analyze the \textit{dog adoption} topic (\cref{fig:dog_emo_sankey}), a common topic discussed across X, Facebook and TikTok.
All three platforms show predominantly positive emotions, but differ in how these emotions translate into engagement.
Facebook is with more caring and love, which then lead to Level 1 interactions, which is stable and affiliative forms of reciprocity.
TikTok is dominated by excitement, producing Level 0 (view) responses of short-lived, attention-driven engagement.
X, shows moderate optimism and curiosity, which \textsc{Dreams} converts into a balanced distribution of Level 1 and Level 2 interactions, matching with its conversational and broadcast hybrid design.
These findings suggest that even under the same topic and shared emotional valence, platform architecture mediates how emotional expressions convert into engagement, aligning with the notion of emotional currency in platform-specific exchange markets.

\textbf{Different emotions drive distinct engagement depths, even on the same platform.}
Within Facebook (2022) (\cref{fig:cc_emo_sankey}), three climate change narratives differ in how emotional tone translates into user engagement. 
Climate-Anxious persona, characterized by fear and sadness, generate higher probabilities of Level 1 engagement.
Climate-Skeptical persona, dominated by disapproval and anger, produces more commenting as deliberative or oppositional exchange.
Pro-Climate narratives, rich in caring and optimism, have a blend of Level 1--3 (like/share/comment) engagements, showing collective reinforcement within the community.
The comparatively lower Level 0 predictions may result from limited supervision in training data for passive engagement signals for Facebook.

\subsection{Ablation Studies}
We examine how each component of \textsc{Dreams} contributes to performance using ablation experiments on both Mamba and Transformer variants.
\cref{tab:detailed_results} reports misinformation engagement performance after removing the memory module (mm), FiLM layer (F), belief state (bel), and disentangled representations (dis). \cref{tab:new_posts_mape} reports performance on forecasting misinformation post volumes.

\textbf{Disentangled representations make the largest  gains.} 
For \textsc{Dreams}-Mamba, removing disentangled representations causes the largest performance drop, leads to MAPE goes up by 39.4\% (from $0.190$ to $0.265$ for 7-day forecasts).
FiLM layers follow with 18.2\% degradation ($0.225$), then memory modules (10.6\%, $0.211$), and belief states (5.8\%, $0.201$).
This pattern shows that the state-space model depends strongly on disentangled features to control information flow across heterogeneous platforms.

\textsc{Dreams}-Transformer shows a slightly different pattern: disentangled representations remain the most important (34.3\% drop, $0.227$ to $0.305$), but belief states contribute more (29.5\%, $0.294$), followed by FiLM (17.2\%, $0.266$) and memory modules (10.8\%, $0.252$).
The consistent dominance of disentanglement in both models confirms that separating platform-specific dynamics from general temporal patterns is essential for cross-platform engagement prediction~\cite{lindstrom2021computational}
For post-volume prediction (\cref{tab:new_posts_mape}), removing disentangled representations or FiLM layers leads to the largest increase in error across both variants, indicating that content disentanglement and platform adaptation are essential.
In contrast, removing memory modules or belief states results in only minor degradation, suggesting that post volumes are governed more by structural platform dynamics than by sequential context.

\textbf{Engagement behavior is short-term and reward-driven.}  
Memory modules show the smallest overall effect, suggesting that engagement behavior depends more on immediate reward signals than on long-term history---consistent with evidence that viral interactions follow rapid ``peak-and-decay'' dynamics~\cite{sangiorgio2025evaluating}.
Interestingly, the higher importance of belief states in Transformer-based (29.5\%) compared to D-Mamba (5.8\%) may indicate that state-space recurrence captures user expectation patterns, while Transformers require explicit belief modeling.

\section{Conclusion}
We present \textsc{Dreams}, a framework for cross-platform misinformation engagement prediction that combines disentangled representation learning, dual-timescale memory, and platform-adaptive modulation.
Grounded in Social Exchange Theory (SET), \textsc{Dreams} models online engagement as a dynamic exchange process---where users' interactions reflect the flow of social value over time and across platforms.
It shows how formal social theories can shape the inductive biases of deep temporal models, leading to architectures that are socially grounded, data-efficient, and adaptable to evolving online ecosystems.
Future work could extend \textsc{Dreams} to causal intervention modeling or multi-agent formulations of social reciprocity.

\section{Acknowledgments}
This research was supported by the Advanced Strategic Capabilities Accelerator (ASCA), the Defence Science and Technology Group, the Defence Innovation Network, and the Australian Academy of Science.

\bibliographystyle{ACM-Reference-Format}
\bibliography{dreams,bds-papers}


\appendix

\newpage

\begin{figure}[b]
   \centering
   \includegraphics[width=\linewidth]{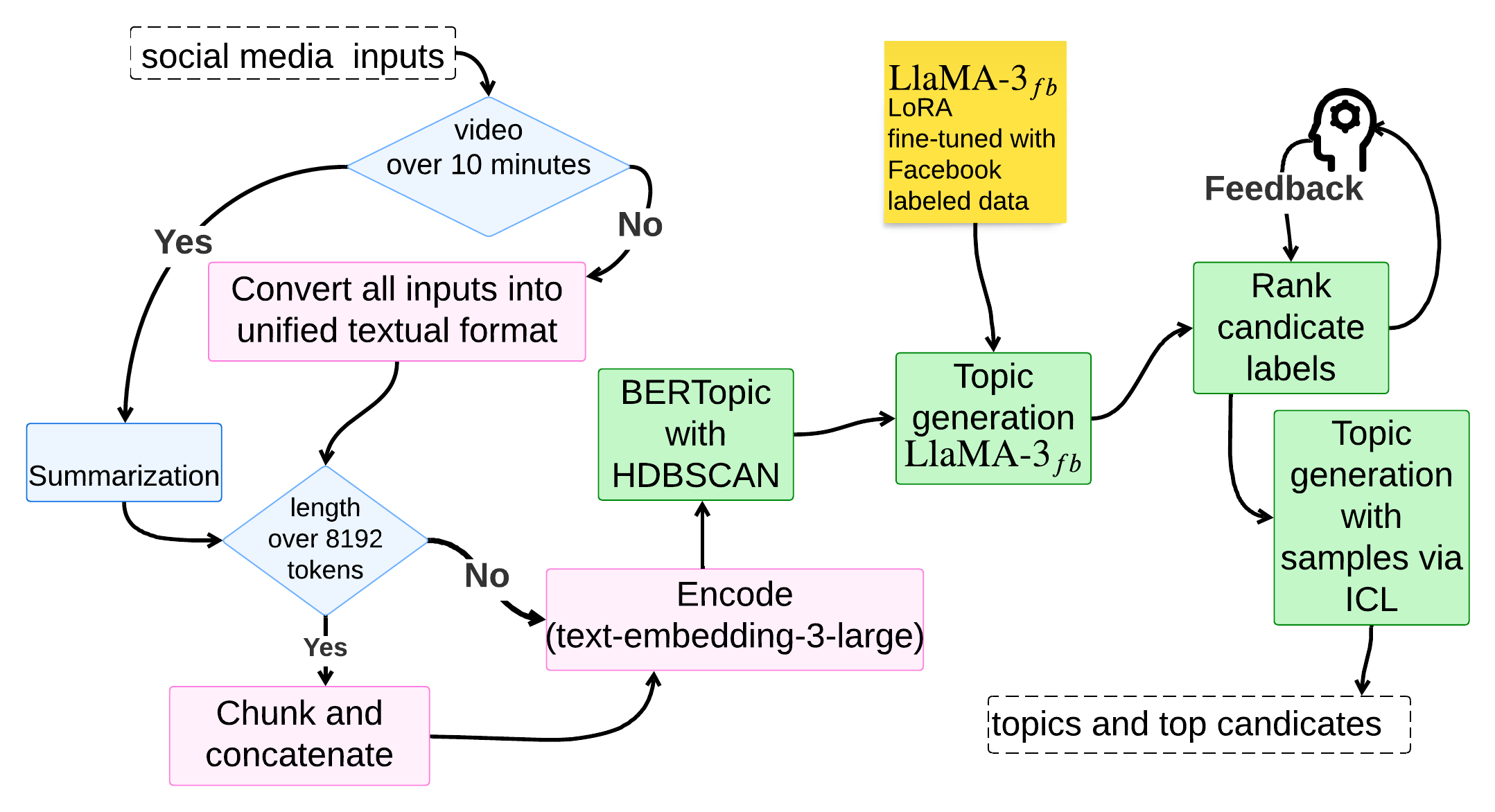}
   \caption{Cross-platform topic and opinion labeling pipeline in \textsc{Dreams}.
      (1) text and video inputs are standardized;
      (2) long videos ($>$10 minutes) are summarized to retain opinion-level content;
      (3) transcripts and metadata tags are concatenated into unified text;
      (4) text is encoded using \texttt{text-embedding-3-large} (3{,}072-d) and L2-normalized;
      (5) semantically similar posts are clustered with \texttt{BERTopic} (HDBSCAN);
      (6) opinion labels are generated using \texttt{LLaMA-3} (LoRA fine-tuned on Facebook-labeled data),
      (7) labels are ranked and validated by human experts;
      and (8) verified examples are incorporated for iterative prompt refinement via in-context learning.
   }
   \label{fig:topic_labeling_pipeline}
\end{figure}

\begin{figure*}[t]
   \centering
   \begin{lstlisting}[style=promptstyle]
<|begin_of_text|><|start_header_id|>system<|end_header_id|>
You are an expert annotator of social and political discourse. 
Given representative posts from one semantic cluster, generate several concise opinion labels (5-10 words each) that describe the dominant stance or opinion expressed. 
For each label, identify a supporting candidate example from the input posts that best illustrates the label's meaning. 
After generating all labels, rank the (label, supporting example) pairs by their estimated relevance to the overall cluster content, considering semantic similarity and frequency of occurrence.
<|eot_id|>
<|start_header_id|>user<|end_header_id|>
[Representative Cluster Samples: Post 1, Post 2, Post 3, ...]
<|eot_id|>
<|start_header_id|>assistant<|end_header_id|>
1. [Opinion Label A]  
   Supporting Candidate: [Excerpt from Post X]  
   Relevance Score: [0-1]

2. [Opinion Label B]  
   Supporting Candidate: [Excerpt from Post Y]  
   Relevance Score: [0-1]

3. [Opinion Label C]  
   Supporting Candidate: [Excerpt from Post Z]  
   Relevance Score: [0-1]
<|eot_id|>
\end{lstlisting}
   \caption{Label generation and ranked candidates prompt template.}
   \label{fig:narrative_prompt}
\end{figure*}

\begin{figure}[t]
   \centering
   \begin{lstlisting}[style=promptstyle]
<|begin_of_text|><|start_header_id|>system<|end_header_id|>
You are an expert content summarizer for social media discourse.
Given a long transcript, produce a concise paragraph (150-200 words) that preserves the core opinion, positions/opinions, and emotional tone.
Remove fillers and redundancies; keep topic continuity and key claims.
<|eot_id|>
<|start_header_id|>user<|end_header_id|>
[Video Transcript + Tags/Hashtags]
<|eot_id|>
<|start_header_id|>assistant<|end_header_id|>
[Concise opinion-level summary]
<|eot_id|>
\end{lstlisting}
   \caption{Summarization prompt template.}
   \label{fig:summarization_prompt}
\end{figure}

\section{Topic Modeling Pipeline}
\label{app:topic_label_pipeline}
For non-Facebook platforms, we construct topic-level annotations through a semi-automated pipeline that integrates multimodal transcription, text embedding, semantic clustering, generative labeling, and expert-guided refinement. 
The pipeline integrates multimodal transcription, large-scale text embedding, semantic clustering, generative opinion labeling, and expert verification to produce consistent topic and opinion annotations across heterogeneous social media platforms. 
An overview is shown in \cref{fig:topic_labeling_pipeline}.

\paragraph{Input Standardization and Preprocessing}
The pipeline first standardizes input across text- and video-based platforms.  
For video data (e.g., Bilibili and TikTok), we extract or generate textual transcriptions as follows:
\begin{itemize}
    \item \textbf{Bilibili:} If an uploaded transcript is available, we use the provided caption text. Otherwise, we generate an automatic speech recognition (ASR) transcript using the \texttt{Whisper} API\footnote{\url{https://platform.openai.com/docs/guides/speech-to-text}}. Video lengths range from 1 to 30 minutes. For videos exceeding 10 minutes, we apply an LLM-based summarization step (as in \cref{fig:summarization_prompt}) to condense the transcript while preserving opinion semantics and key topics.
    \item \textbf{TikTok:} Videos are typically under 60 seconds. We use available captions or generate transcripts using \texttt{Whisper}. 
\end{itemize}
For both platforms, hashtags and user-provided tags are concatenated with transcripts to form a unified textual input.  

\paragraph{Text Embedding and Representation}
All inputs are converted into dense vector representations using the \texttt{text-embedding-3-large} model, producing 3{,}072-dimensional embeddings for each document.  
Embeddings are then normalized using L2 normalization\footnote{\url{https://docs.pytorch.org/docs/stable/generated/torch.nn.functional.normalize.html}} to ensure uniform magnitude before clustering.  

\paragraph{Semantic Clustering with \texttt{BERTopic}}
To discover opinion clusters, we apply the \texttt{BERTopic} framework\footnote{\url{https://huggingface.co/docs/hub/en/bertopic}}, which combines the embeddings and hierarchical density-based clustering (HDBSCAN).  
Each cluster represents a semantically coherent group of posts or transcripts that share similar linguistic or thematic characteristics.  
We assess cluster coherence using cosine similarity between normalized embeddings.

\paragraph{Post-hoc opinion Labeling with \texttt{LLaMA-3}}
We assign interpretable opinion labels to each cluster using a generative large language model. 
We use \texttt{LLaMA-3}\footnote{\url{https://huggingface.co/meta-llama/Meta-Llama-3-70B-Instruct}} fine-tuned with LoRA on the manually labeled Facebook subset from~\cite{kong2022slipping}.
The LoRA adaptation approach has proven effective for various text classification tasks, including multilingual content moderation and detection \cite{Tian2025Exist}, where hierarchical fine-tuning with low-rank adaptation enables efficient transfer learning across languages and platforms. 
The model then generates candidate opinion labels summarizing the dominant opinion within that cluster.  
In addition, \texttt{LLaMA-3} ranks the support candidates by estimated relevance, incorporating cluster-level similarity scores as soft priors, as shown in \cref{fig:narrative_prompt}.

\paragraph{Human-in-the-Loop Verification and Iterative Refinement}
Generated labels go under expert verification. Human annotators review candidate labels, top-ranked samples, and intra-cluster similarity scores, providing binary relevance feedback.  
Based on expert feedback, prompts are updated through in-context learning, adding verified examples and excluding rejected ones. 


\section{Dataset}
\label{app:dataset_more}
\begin{table*}[t]
\centering
\caption{\textbf{Cross-platform misinformation engagement dataset.}
Each platform varies in language coverage, time span, and engagement actions.
L0--L3 correspond to platform-specific interaction types reflecting increasing effort.
EN = English, ZH = Chinese, RU = Russian.}
\label{tab:dataset_more}
\resizebox{\textwidth}{!}{
\begin{tabular}{lcccccccc}
\toprule
\textbf{Platform} & \textbf{Language(s)} & \textbf{Time Span} & \textbf{Modality} & \textbf{L0 (View)} & \textbf{L1 (Like)} & \textbf{L2 (Share/Repost)} & \textbf{L3 (Comment/Reply)} \\
\midrule
Facebook & EN & 2021--2024 & text,image & -- & Like/Emoji & Share & Comment \\
Instagram & EN & 2022--2024 & text,image & -- & Like & -- & Comment \\
X (Twitter) & EN, & 2021--2023 & text & -- & Like & Retweet & Reply \\
Telegram & RU, EN & 2021--2024 & text & View & -- & Forward & -- \\
Bilibili & ZH & 2021--2025 & text,video & Play & Like & Repost & Comment \\
Bluesky & EN & 2024--2025 & text & -- & Like & Repost & Reply \\
TikTok & EN & 2024--2025 & text,video & Play & Like & Share & -- \\
\bottomrule
\end{tabular}}
\end{table*}

\Cref{tab:dataset_more} summarizes the cross-platform misinformation engagement dataset used in this study with more details.
Each platform differs in language coverage, temporal range, and interaction structure, with engagement levels (L0--L3) mapped to their native action names (e.g.,\ \textit{View}, \textit{Like}, \textit{Share}, \textit{Comment}).

The dataset's temporal structure and opinion-level organization enable not only engagement prediction but also potential extensions to causal analysis of influence \cite{Tian2025Causal} and detection of coordinated information operations \cite{Tian2025}. The multi-platform coverage allows for comparative studies of content moderation effectiveness across different policy environments \cite{Schneider2023}, while the fine-grained engagement hierarchy captures the spectrum of user investment levels that emerge from platform-specific social exchange dynamics \cite{Schneider2025}.

%


\end{document}